\documentclass[%
 reprint,
 superscriptaddress,
 amsmath,amssymb,
 aps,
]{revtex4-2}

\usepackage{graphicx}  
\usepackage{dcolumn}   
\usepackage{bm}        
\usepackage{comment}

\newcommand{\newfigureWIDEtoph}[5]{
  \begin{figure*}[!ht]
    \centering
    \includegraphics[width=#5]{#1}
    \caption[#2]{\textbf{#2.} #3}
    \label{#4}
  \end{figure*}
}

\begin{document}

\preprint{APS/123-QED}

\title[A model of replicating coupled oscillators...]
{A model of replicating coupled oscillators generates naturally occurring cell networks}

\author{Matthew Smart}
\email{msmart@flatironinstitute.org}
\affiliation{Center for Computational Biology, Flatiron Institute, New York, NY}

\author{Stanislav Y. Shvartsman}
\email{stas@princeton.edu}
\affiliation{Center for Computational Biology, Flatiron Institute, New York, NY}
\affiliation{Department of Molecular Biology, Princeton University, Princeton, NJ}
\affiliation{Lewis-Sigler Institute for Integrative Genomics, Princeton University, Princeton, NJ}

\author{Hayden Nunley}
\email{hnunley@flatironinstitute.org}
\affiliation{Center for Computational Biology, Flatiron Institute, New York, NY}

\begin{abstract}
When a founder cell and its progeny divide with incomplete cytokinesis, a network forms in which each intercellular bridge corresponds to a past mitotic event. 
Networks built in this manner are required for gamete production in many animals, and different species have evolved very different final network topologies.
While mechanisms regulating network assembly have been identified in particular organisms, we lack a quantitative framework to understand network assembly and inter-species variability.  
Motivated by cell networks responsible for oocyte production in invertebrates, where the final topology is typically invariant within each species, we devise a mathematical model for generating cell networks: each node is an oscillator, and after a full cycle, the node produces a daughter to which it remains connected. 
These cell cycle oscillations on the nodes are transient and coupled via diffusion over the network's edges. 
By variation of three biologically motivated parameters, our model generates nearly all such networks currently reported across invertebrates. 
Furthermore, small parameter variations can rationalize cases of within-species variation. Because cell networks outside of the ovary often form less deterministically, we propose generalizations of our model to account for different sources of stochasticity.
\end{abstract}

\maketitle

A network of cells joined through intercellular bridges (ICBs) is formed when a founder cell and its progeny undergo incomplete divisions 
\cite{CHAIGNE2022R385}. Examples of networks built by this process include colonies of choanoflagellates \cite{FAIRCLOUGH2010R875, Larson2020_choano} -- the closest living relatives of animals -- and both male and female germline cysts \cite{fawcett_occurrence_1959, fawcett_intercellular_1961, Haglund2011, lu_stay_2017}. The first example demonstrates this process's potential role in the evolution of multicellularity, while the second illustrates its importance in animal reproduction. In both cases the networks are typically ``small", having fewer than one hundred cells. Despite the ubiquity of small cell networks in nature, a theory for their generation via coupled, transient oscillations is lacking. 

We turn to perhaps the best-studied class of cellular networks: the ovarian germline cysts of invertebrates (Fig. \ref{fig:dros_oocyte_plus_species}), which are well suited for mathematical modeling because their final topology is often stereotyped within a given species, but variable across species. These networks form inside compartments of the ovary which function like assembly lines. At one end of each compartment, a germline stem cell generates a daughter cell called a cystoblast. As the cystoblast moves away from the germline stem cell, the cystoblast and its progeny divide incompletely, resulting in a network of cells (germline cyst) connected by ICBs. A subset of these cells, typically only one, become oocytes. The resulting ovarian germline cysts can be divided into four categories based on topology (Fig. \ref{fig:dros_oocyte_plus_species}c-f). The first class of cysts is linear chains of various lengths. The cysts in the second class can be formed by $n$ rounds of synchronous divisions with branching, resulting in $2^{n}$ cells arranged in a topology called ``maximally branched''. Cysts in the third class contain $2^{n}$ cells, but the cells form linear chains connected to each other via a branched core. The last class contains branched cysts with non-$2^{n}$ number of cells. 

With this diversity of natural forms in mind, and building on experimental studies in the fruit fly \emph{Drosophila melanogaster}, we devise a model by which such networks are assembled step-by-step through the mitotic addition of nodes and edges. 
The model deterministically maps a single-cell initial condition to a final network, the topology of which depends on the value of several key parameters such as the strength of cell-cell coupling and the asymmetry of cell division. 
We demonstrate that the model is able to generate nearly all reported invertebrate cysts and identify their relative locations in the space of model parameters.
This in turn allows us to characterize neighboring regions where a small change in parameters causes one final network to be converted to another. 
We discuss how our framework could guide genetic manipulations of network formation, while also providing insight into the origin of intra-species network variability. 
Finally, we discuss generalizations based on alternative rules for incorporating new nodes \cite{JIALSOUS2021, diegmiller_fusome_2023}, which will expand the space of possible graphs that can be generated through a finite sequence of incomplete divisions.

\newfigureWIDEtoph{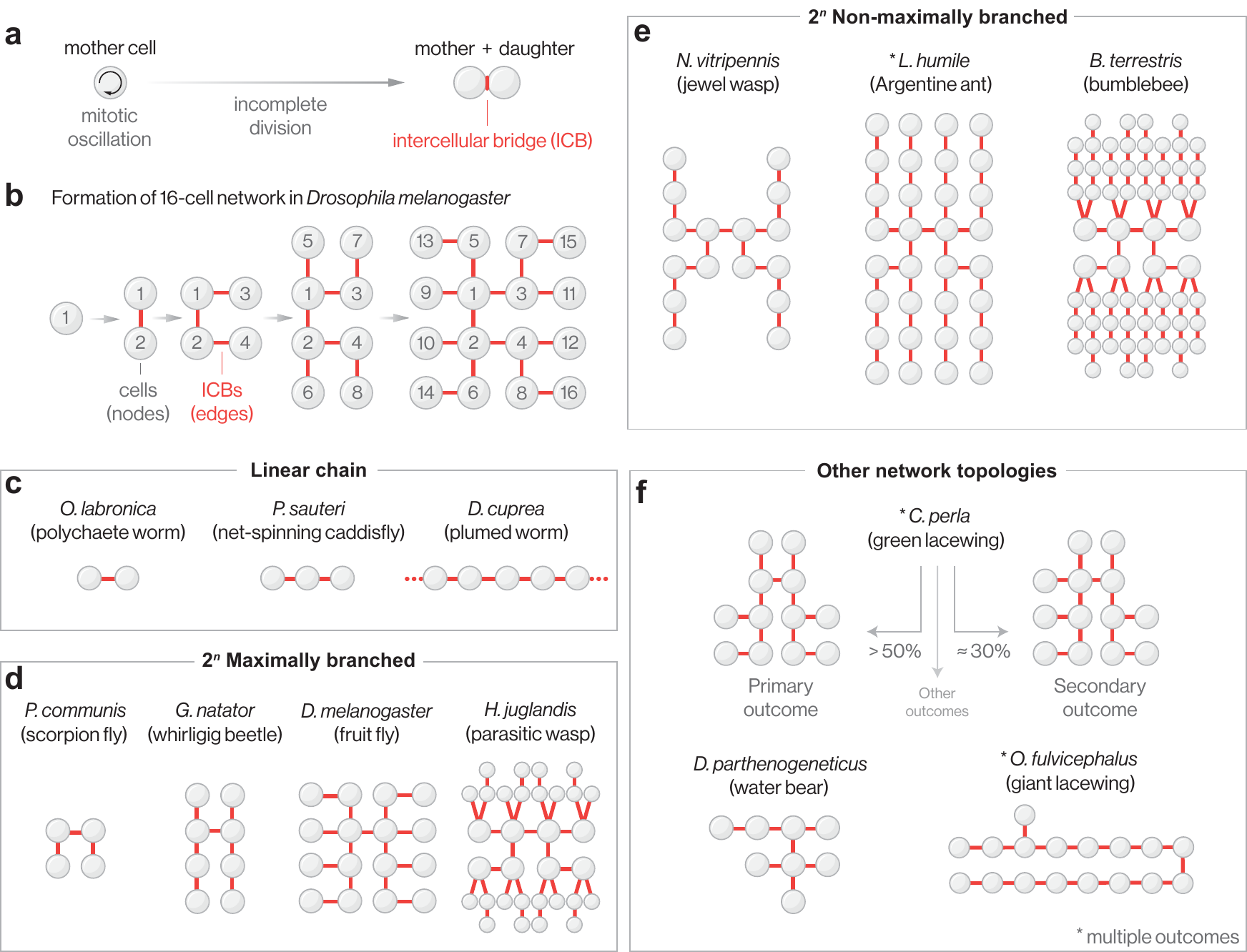}
  {The topology of germline cysts -- biological networks formed by coupled transient oscillations -- exhibit significant inter-species variation}
  {\textbf{a}, Cell cycle oscillations in a mother cell initiate an incomplete division event that results in two cells connected by an intercellular bridge (ICB). 
\textbf{b}, Formation of a cell network responsible for producing oocyte(s) (known as a germline cyst) in female \emph{D. melanogaster}. A cystoblast undergoes four serial, synchronous divisions to generate a 2-, 4-, 8-, and finally a 16-cell network (see Fig. S1 for details). This final topology is invariant in wild-type \emph{D. melanogaster}. Nodes correspond to cells and edges to ICBs. Nodes are labeled according to division order (grouped by synchrony: 1, 2, 3-4, 5-8, 9-16).
\textbf{c-f}, Zoology of germline cyst topologies. For each of the listed example species, the final network topology is invariant, unless indicated otherwise. Topologies are categorized as linear (\textbf{c}), ``$2^n$ maximally branched" (\textbf{d}), ``$2^n$ non-maximally branched" (\textbf{e}), or ``other" (\textbf{f}). Of the species shown here, \emph{L. humile}, \emph{C. perla}, \emph{O. fulvicephalus} exhibit intraspecies network variation.}
{fig:dros_oocyte_plus_species}{0.9\linewidth}

\section*{\label{sec:results}Results}

\subsection*{\label{sec:model}Model of coupled oscillators that replicate through incomplete division}

Below we present a model of multicellular network assembly where each node is a cell cycle oscillator. 
In brief, each full oscillation of a node (mother cell) generates another oscillator (daughter cell) to which it remains diffusively coupled \cite{Kouvaris2012} via a network edge (ICB). 
The network reaches some final topology once all cells have stopped dividing. 
The model is outlined in Fig. \ref{fig:model_fig}, and Table S1 provides an overview of terms and expressions. 
Additional details and modeling assumptions are in Methods and SI. 

\subsubsection*{\label{sec:model_sc}Dynamics of a single cell oscillator}

Each cell is treated as a dynamical system with state $\mathbf{x}\in\mathbb{R}^N$ that evolves according to

\begin{equation}
\label{eq1}
\frac{d\mathbf{x}}{dt}=\mathbf{f(x)}+\mathbf{h}(t)
\end{equation}
where $\mathbf{f(x)}$ encodes the intracellular dynamics and $\mathbf{h}(t)$ controls when oscillations occur.

We consider a minimal form for $\mathbf{f(x)}$ with $N=2$ dynamic variables $x,y$ motivated by the known biology of cell cycle components \cite{QiongYang2013, FerrellYang2011, MATTINGLY2017743, DENEKE2016399, NOVAK1993101}. 
The simplest setting involves oscillations of activated ($x$) and total ($y$) concentrations of mitotic cyclin-CDK complexes. 
In an abstracted form of the conventional cell cycle models, their dynamics are 
\begin{equation}
\label{eq2}
\begin{aligned}
\epsilon \frac{dx}{dt} &= y-g(x) \\ 
         \frac{dy}{dt} &= z-x-\gamma y
\end{aligned}
\end{equation}
where $g(x)$ is a piecewise linear function encapsulating the nonlinear regulation of $x$ (see Methods). 
Motivated by relaxation oscillations in embryonic cell cycles \cite{QiongYang2013, pomerening_building_2003, POMERENING2005565}, we assume $0<\epsilon\ll 1$ (i.e. $x$ evolves fast relative to $y$). To preclude bistability, we further assume that $0<\gamma<\frac{1}{2}$. 

The parameter $z$ controls whether the above system will tend to a unique stable fixed point or a limit cycle by shifting where the nullclines for $x$ and $y$ intersect.
Intersection on the left or right segments of $g(x)$ (which defines the $x$-nullcline) implies a stable fixed point, whereas
intersection on the middle segment leads to limit-cycle oscillations (see Fig. \ref{fig:model_fig}a and Methods). 

Accordingly, the initiation and termination of oscillations can be controlled by moving the system into and then out of the oscillatory regime. 
Motivated by studies of \emph{D. melanogaster} germline cysts \cite{MathieuHuynh2017, Insco2009, ShanmingChaoyi2017, McKearin01121990, McKearinOhlstein1995}, we consider a time-dependent regulation $\mathbf{h}(t)$ in Eq. (\ref{eq1}) with a ``pulsed" form.
To implement this, we assume for simplicity that $z(t)$ is a triangular pulse (see Fig. \ref{fig:model_fig}a,b and Methods). 
Because $z(t)$ positively regulates $y$, it promotes mitotic entry then exit by translating the $y$-nullcline into and out of the oscillatory regime.
The profile of $z(t)$ will therefore strongly affect the number of transient oscillations of both $x$ and $y$ and, thus, the finite growth of the network (outlined later).

\newfigureWIDEtoph{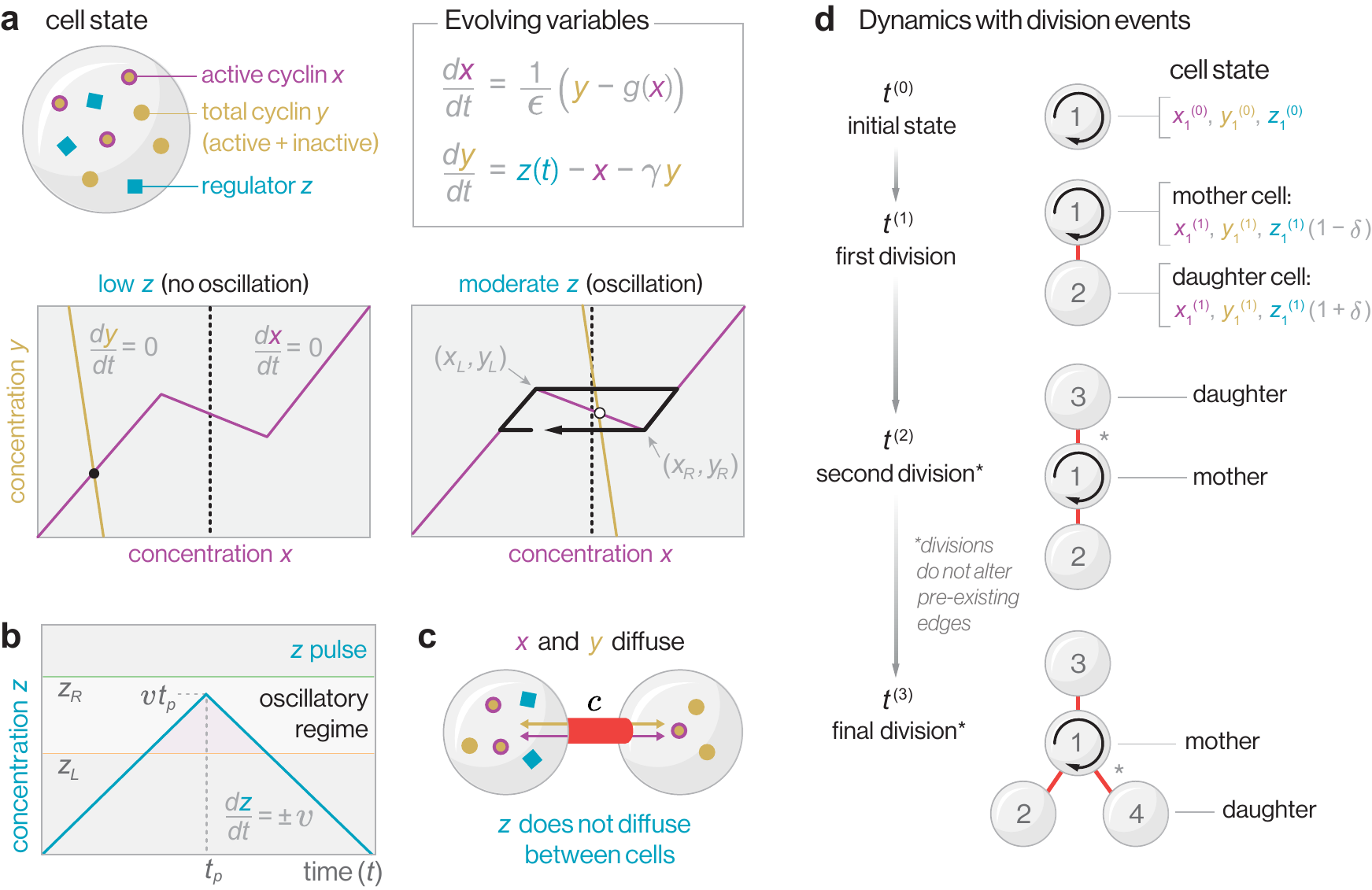}
  {Model overview: building small networks by oscillations on nodes}
  {\textbf{a}, 
Cell cycles depend on the oscillating concentrations of active cyclin $x$ and total cyclin $y$, modeled as a relaxation oscillator. $\epsilon$ controls the relative timescale of $x$ and $y$ dynamics.
The level of a regulatory variable $z$ controls whether oscillations occur. 
Low levels of $z$ prevent oscillations, while moderate levels of $z$ cause oscillations in $x$ and $y$.
A full cycle of crossings through the black dashed line triggers a division.
\textbf{b}, 
A pulse in $z$ allows the system to enter and exit the oscillator regime (bounded by solid orange and green lines) by translating the $y$-nullcline. 
We use a symmetric triangle pulse with a slope of magnitude $v$ and duration $2 t_{p}$.
Pink triangle: the range of ($t$, $z$) for which the oscillator cycles.
\textbf{c}, 
In a network of coupled oscillators, the oscillator variables $x$ and $y$ both diffuse over network edges (representing ICBs) with rate $c$.
\textbf{d},  
Network dynamics starting from a single cell with initial condition $x_1^{(0)}, y_1^{(0)}, z_1^{(0)}$. 
New nodes are generated each time a cell completes an oscillation. 
Cell divisions do not alter pre-existing edges.
Following a division, the levels of the oscillator variables $x$ and $y$ are copied from the pre-division mother to the post-division mother and daughter, whereas the cell cycle regulator $z$ may be distributed unequally, as determined by $\delta$. In this example, three oscillations in a founding cell -- but no oscillations in its daughters -- generates a four-cell star network. }
  {fig:model_fig}{1.0\linewidth}

\subsubsection*{\label{sec:model_network}Dynamics of a fixed network of coupled cells}

The state of a network of $M$ cells depends on each cell's internal state $(x_a, y_a)$ and regulation $z_a$, where $a \in \{1,\dots, M\}$ indexes distinct cells.
The adjacency matrix $\mathbf{A}$ -- which is symmetric, unweighted, and of size $M\times M$ -- describes the presence ($A_{ab}=1$) or absence ($A_{ab}=0$) of an ICB between two cells (Fig. \ref{fig:dros_oocyte_plus_species}). 
Since ICBs facilitate diffusion of components between adjacent cells during network formation \cite{SnappIida2004}, we allow the oscillator variables $x_a, y_a$ to diffuse with rate $c$ over the ICB network as a minimal form of cell-cell coupling. 
The multicellular dynamics are then
\begin{equation}
\label{eq:multicell}
\begin{split}
 \frac{d x_a}{dt}  = 
    \frac{1}{\epsilon}(y_a - g(x_a)) 
    &+ c \sum_{b=1}^M A_{ab} (x_b - x_a) \\
\frac{d y_a}{dt}  = 
    z_a - x_a - \gamma y_a  
    &+ c \sum_{b=1}^M A_{ab} (y_b - y_a)
\end{split}
\end{equation}
which accounts for the interplay between intracellular regulation (Eq. (\ref{eq2})) and intercellular diffusion with rate $c$.
Note that when $c=0$, each of the $M$ cells evolves independently according to Eq. (\ref{eq2}). 
For large $c$, one expects synchronized oscillations of all $M$ cells. 
For intermediate $c$, non-intuitive collective effects can arise because the extent of oscillator asynchrony depends on $\mathbf{A}$, while the structure of $\mathbf{A}$ itself is modified by the oscillations.

\subsubsection*{\label{sec:model_network}Incomplete division events enable network formation}

Oscillations of cell cycle components $x_a$, $y_a$ drive the incomplete division of a mother cell $a$ into two cells \cite{TYSON2008R759, HinnantAbles2020, HINNANT2017118} that remain physically connected by an ICB. In our model, a cellular division event occurs whenever an indicator variable for the oscillations (in our case $x$, see Methods) cyclically crosses a threshold (vertical dashed lines in Fig. \ref{fig:model_fig}a). 
After the criterion for a division event is satisfied, two network attributes must be specified: the expanded adjacency matrix and the post-division cell states. 

First, to update the adjacency matrix at each division, we assume that newly generated cells form a new branch (i.e. a new node and edge) without altering existing edges (Fig. \ref{fig:model_fig}d). 
This assumption is based on studies in female germline cysts of \emph{D. melanogaster} and other insects \cite{DIEGMILLERREVIEW, McGrailHays1997, StortoKing1989, TELFER1975223} (see Discussion for alternative choices \cite{JIALSOUS2021, diegmiller_fusome_2023, tworzydlo_architecture_2019}). 
The ``post-division mother" occupies the mother's original position in the network, while the ``post-division daughter" forms the new branch.

Second, to relate the state of the pre-division mother $a$ to the states of the post-division mother $a^\prime$ and daughter $b^\prime$, we specify the maps $\mathbf{x}_a\rightarrow(\mathbf{x}_{a^\prime},\mathbf{x}_{b^\prime})$ and $z_{a}\rightarrow(z_{a^\prime},z_{b^\prime})$. 
When both maps are symmetric, the model only generates ``maximally branched'' networks of $2^{n}$ cells (Fig. \ref{fig:dros_oocyte_plus_species}d). 
As a minimal implementation of asymmetric division \cite{de_cuevas_morphogenesis_1998, McGrailHays1997, diegmiller_fusome_2023}, we assume copying of the oscillator variables
$\mathbf{x}_a\rightarrow(\mathbf{x}_{a},\mathbf{x}_{a})$, 
but allow the regulatory variable $z$ to be asymmetrically inherited as
$z_{a}\rightarrow((1 - \delta)z_{a},(1+\delta)z_{a})$ 
with $\vert \delta\vert < 1$ (Fig. \ref{fig:model_fig}d).
A non-zero $\delta$ generates $z$ differences between progeny and, thus, introduces heterogeneity in the network of oscillators by modifying their respective periods $T(z)$ (see Methods, SI). 

\subsection*{\label{sec:results_lacewing}Asymmetric division events produce natural networks}

To illustrate how the model can generate naturally occurring networks from a single cell, Fig. \ref{fig:buildlacewing} displays the model-predicted sequence of divisions for a parameter set where the final network matches the main variant of female \emph{Chrysopa perla} (green lacewing, see Fig. \ref{fig:dros_oocyte_plus_species}f) germline cysts \cite{rousset_formation_1978}. At time $t = 0$, there is one cell (cell 1) for which the state, $x$ and $y$, resides at a unique stable fixed point. As the pulsed variable $z$ increases, this fixed point shifts until becoming destabilized, giving way to limit-cycle oscillations (Fig. \ref{fig:model_fig}a). After one cycle (Fig. \ref{fig:buildlacewing}b), cell 1 produces cell 2 (Fig. \ref{fig:buildlacewing}a), while retaining more $z$ than cell 2. Both complete their next oscillation nearly synchronously, generating a linear 4-cell network. After this, the $z$ variable declines. Cells 3-4 divide nearly synchronously, while cells 1-2 do not divide (Fig. \ref{fig:buildlacewing}a,b). The result is a linear 6-cell network in which each cell undergoes one final division nearly synchronously. 

Not only does the generated final network (Fig. \ref{fig:buildlacewing}a) agree with the primary \emph{C. perla} variant, but also the division history matches the order of events inferred from fixed samples \cite{rousset_formation_1978}, with cells 1-2 skipping the third round of mitosis before dividing in the final round. 
In our model, this ``skipping" phenomenon occurs due to a transiently desynchronizing effect of moderate diffusive coupling (explored below). Since the $z$ pulse limits the time window over which oscillations can occur, transient delays have a permanent effect on the network's structure. 

Finally, we note that other regions of parameter space (with much stronger cell-cell coupling) also produce this primary lacewing variant. However, they have qualitatively different network formation trajectories (see Fig. S2) and, importantly, do not neighbor secondary and tertiary lacewing variants in parameter space (explored below; see Methods and Fig. S4).


\newfigureWIDEtoph{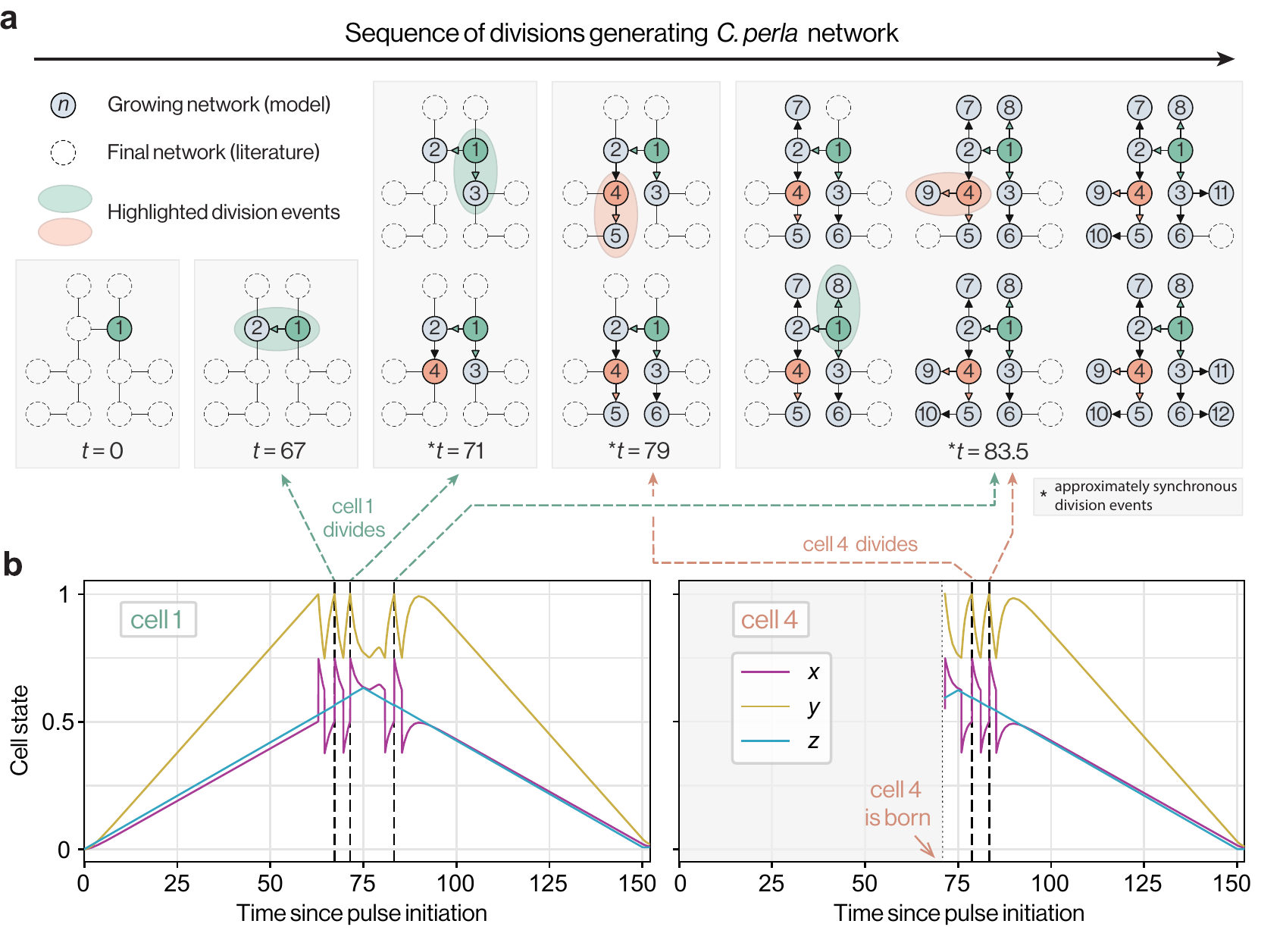}
  {Building a \emph{C. perla} cyst via oscillations on nodes}
  {\textbf{a}, Model-generated sequence of divisions grouped by approximate synchrony. Underneath each step is a skeleton of the most common \emph{C. perla} cyst variant. Division events of cells 1 and 4 are highlighted in green and orange, respectively.
\textbf{b}, Time series of $x$, $y$, and $z$ in the cyst's founder cell (1, left), and another cell (4, right). The trajectory of cell 4 begins at its birth as indicated along the time axis. Dashed lines depict how oscillations in \textbf{b} map to division events of a corresponding cell in \textbf{a}.}
  {fig:buildlacewing}{0.9\linewidth}


\newfigureWIDEtoph{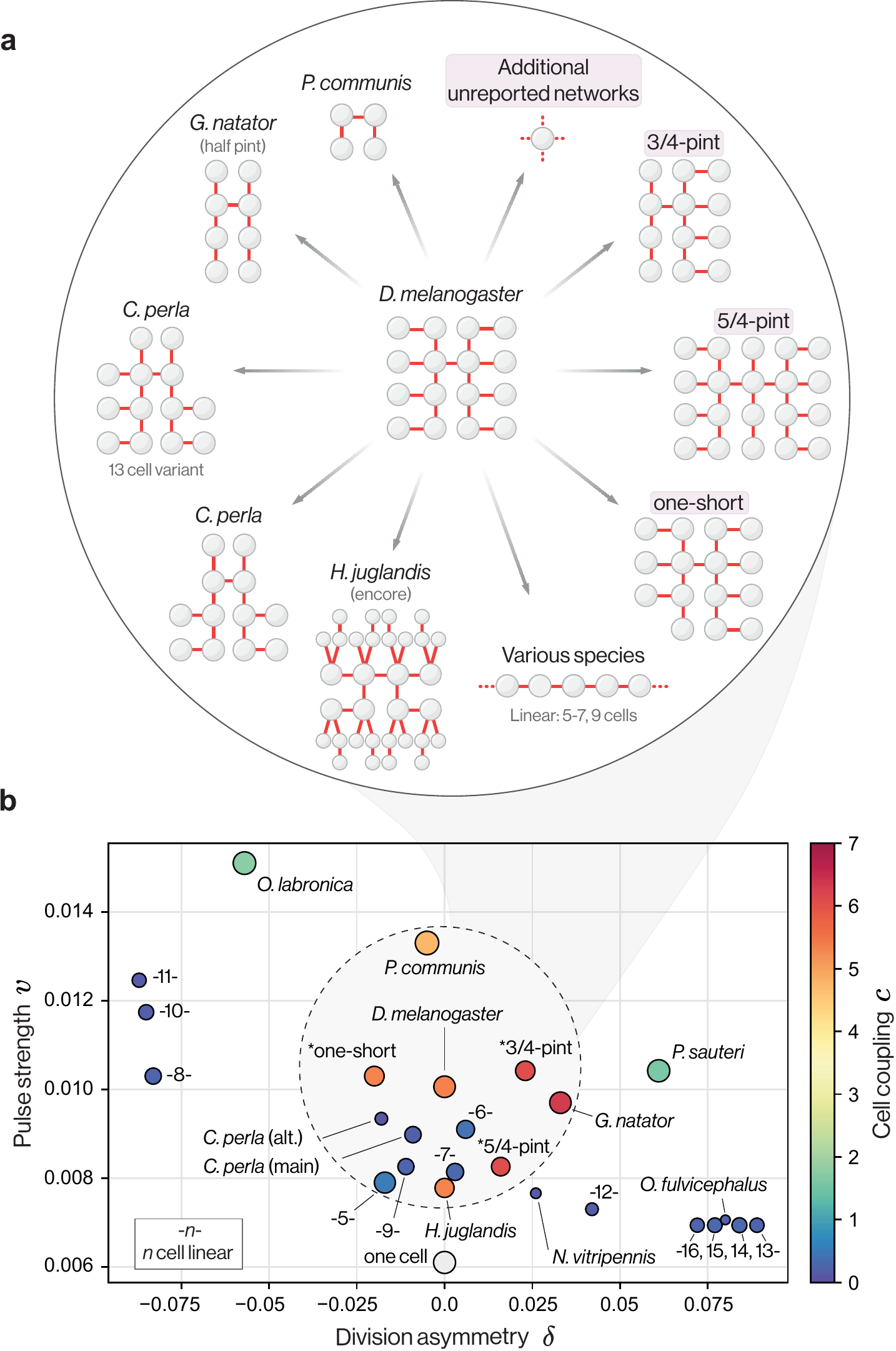}
  {Relating naturally occurring cell networks through changes in model parameters}
  {\textbf{a}, Parameter scans over $c$, $v$, and $\delta$ reveal which final networks are adjacent (gray arrows) to \emph{D. melanogaster}'s cyst in parameter space. 
Pink labels indicate networks that are not reported to occur in nature but border the \emph{D. melanogaster} region over a large range of parameters.
\textbf{b}, 
The centroids of parameter regions that produce networks from Fig. \ref{fig:dros_oocyte_plus_species}c-f are shown as points (see Methods). 
Relative size reflects the fraction of parameter sets that produced the labeled network. 
Numbered points denote linear networks of varying size.
The gray circle centered on \emph{D. melanogaster} highlights the locations of its neighboring networks shown in \textbf{a}.}
  {fig:cubesfigure}{0.6\linewidth}


\newfigureWIDEtoph{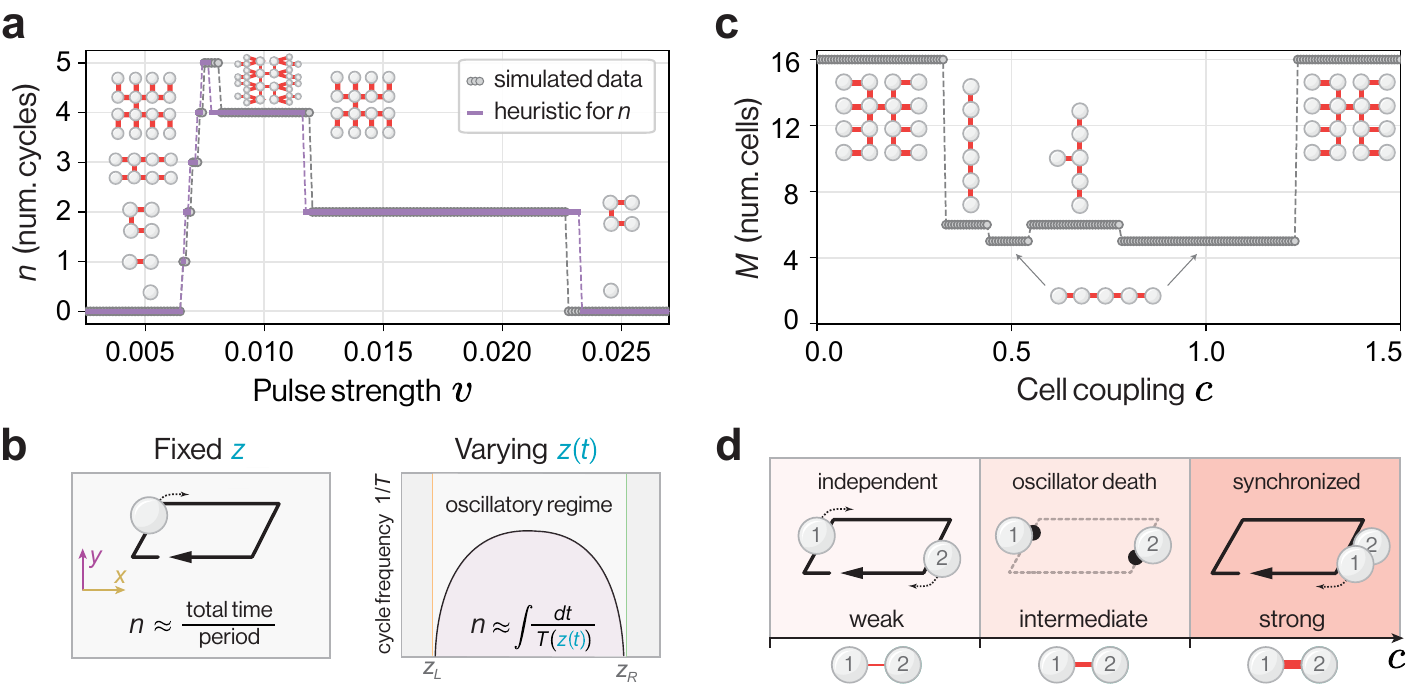}
  {The effect of pulse strength $v$ and cell coupling $c$ on the final network}
  {\textbf{a}, Number of cycles $n$ for varying $v$ with no division asymmetry ($\delta=0$). 
\textbf{b}, Heuristic $n(v)$ for a single cell. Left: Estimate of $n$ for fixed $z$. Right: Estimate of $n$ for time-varying $z(t)$ based on an integral of the cycle frequency over time $t$ (Fig. \ref{fig:model_fig}b and Methods).
\textbf{c}, Final number of cells $M$ in networks generated by weakly positive asymmetric divisions ($0 <\delta \ll 1$). Though \emph{D. melanogaster}'s cyst is generated at weak and strong coupling, intermediate $c$ generates networks with fewer cells and fewer branches.
\textbf{d}, Two-cell schematic for the role of coupling strength. ICB thickness represents strength of cell coupling $c$.}
  {fig:twoparameterfig}{0.8\linewidth}


\subsection*{\label{sec:results_cube}Identifying regions in parameter space which generate naturally occurring networks}

We next studied which of the natural networks (Fig. \ref{fig:dros_oocyte_plus_species}d) the model is able to generate by performing multi-dimensional parameter scans along pulse strength $v$, division asymmetry $\delta$, and diffusion $c$ (see Fig. \ref{fig:cubesfigure} and Methods). 
Each point in parameter space generates a particular final network, in the same manner as Fig. \ref{fig:buildlacewing}.
The resolution of the parameter sweeps allows us to identify contiguous region(s) of parameter space which generate the same final network (Fig. S4). 
This allows us to define distinct parameter regions as being ``neighbors" when their surfaces are in direct contact (at the resolution of our search). 
Given two neighboring regions, small variations in $v$, $c$, or $\delta$ could allow one to generate either of the associated final networks. 

A key example relating one network (the \emph{D. melanogaster} cyst) to its neighbors is shown in Fig. \ref{fig:cubesfigure}a. 
Among the neighboring natural networks, we note that two of them, \emph{Gyrinus natator} (whirligig beetle) and \emph{Habrobracon juglandis} (parasitic wasp), correspond to established genetic perturbations of wild-type \emph{D. melanogaster}: they can be reached by mutations in either \emph{half pint} (16 to 8 cells) \cite{van_buskirk_half_2002} or \emph{encore} (16 to 32 cells) \cite{hawkins_encore_1996,ohlmeyer_encore_2003}.
We also find that small variations in parameters can transform the \emph{D. melanogaster} cyst into linear networks of various lengths, as well as non-$2^{n}$ branched networks, such as the two most common \emph{C. perla} variants (see Fig. S3b).

Besides the reported natural networks from Fig. \ref{fig:dros_oocyte_plus_species}, many as-yet unreported networks neighbor the region of parameter space that generates \emph{D. melanogaster} cysts. 
As examples, Fig. \ref{fig:cubesfigure} shows the three with the largest contact boundaries (see Methods). 
Inspired by the \emph{half pint} name, we name two of these \emph{$\frac{3}{4}$-pint} and \emph{$\frac{5}{4}$-pint} due to loss and gain of four cells, respectively, relative to \emph{D. melanogaster}. 
We term the remaining network \emph{one-short} because it differs from the \emph{D. melanogaster} cyst by one cell.  
Model-predicted division sequences for generating these three networks are depicted in Fig. S5a-c.

Aside from networks directly neighboring \emph{D. melanogaster}, we identify regions for each of the linear cyst structures from 2-16 cells, one of the reported cysts in \emph{Osmylus fulvicephalus} (giant lacewing), and the \emph{Nasonia vitripennis} (jewel wasp) germline cyst (Fig. \ref{fig:cubesfigure}b). The model also produces a collection of linear cysts of larger lengths (not shown in Fig. \ref{fig:cubesfigure}b). 
Though all these networks appear in the parameter space, some, like that of \emph{D. melanogaster} or \emph{Panorpa communis} (scorpionfly), occupy large parameter regions relative to others such as \emph{O. fulvicephalus} or \emph{N. vitripennis}.
Some natural cyst structures -- like that of \emph{Dactylobiotus parthenogeneticus} (water bear), \emph{Linepithema humile} (Argentine ant), or \emph{Bombus terrestris} (buff-tailed bumblebee) -- are not yet identified by this initial search.
In the SI, we describe an extension of the model which enables the generation of both the \emph{D. parthenogeneticus} and \emph{L. humile} networks by allowing the division asymmetry $\delta$ to depend on the number of previous divisions of the mother.

\subsection*{\label{sec:results_parameterdependence}Pulse strength and cell coupling affect network size and structure}

To develop an understanding for the relationships between networks identified above (Fig. \ref{fig:cubesfigure}), we next consider the dependence of the final network on one parameter at a time (Fig. \ref{fig:twoparameterfig}). 
We focus on the pulse strength $v$ (Fig. \ref{fig:twoparameterfig}a) and the coupling strength $c$ (Fig. \ref{fig:twoparameterfig}c) because of their generically non-monotonic effect on network size and their experimental relevance. 

The dependence on $v$ is most easily understood for the case of symmetric divisions ($\delta=0$). 
Because the cells are identical in this case, the model can only generate networks of size $2^{n}$, where $n$ is the number of rounds of division.
Fig. \ref{fig:twoparameterfig}b provides a heuristic explanation of the dependence of $n$ on $v$.
These rely on knowing the period of oscillations $T(z)$, which we estimate in the SI.
When $z$ is fixed, the number of cycles is the ratio of the time spent in the oscillatory region to the period $T(z)$.
For the case of changing $z(t)$ -- and thus a changing cycle period $T(z)$ -- 
the analogous computation depends on the integral of $\frac{1}{T(z(t))}$ with respect to time $t$ (see Methods). 
Fig. \ref{fig:twoparameterfig}a shows the heuristic result and the simulation data: over a narrow range in $v$, $n$ rapidly increases from $n=0$ to $n=5$, thereafter decreasing in a series of plateaus. 
Fig. S3a shows how these plateaus are altered in the case of relatively strong diffusion $c$ and division asymmetry $\delta$.

Cell coupling $c$ has particularly interesting effects on final network structure; while the network topology itself is modified by the oscillations, the extent of oscillator synchrony depends on network topology. 
To understand the role of $c$, consider the case of two coupled oscillators -- each with static $z$ values -- that are initially desynchronized (Fig. \ref{fig:twoparameterfig}d). 
One can show (see SI) that both cells will oscillate for sufficiently weak or strong coupling, but for intermediate values of $c$ they become trapped in a non-oscillatory collective fixed point. 
This effect is commonly referred to as ``oscillator death" \cite{bar-eli_stability_1985, bar-eli_coupling_1984, ermentrout_oscillator_1990} in networks of fixed topology. Although the pulsed bifurcation parameter prevents our networks from experiencing indefinite oscillator death, the effect of transient desynchronization leaves its mark on the final network structure.

Fig. \ref{fig:twoparameterfig}c shows a characteristic signature of the transient desynchronization associated with intermediate coupling. 
At weak coupling, the final network is that of \emph{D. melanogaster} because the division asymmetry is weak enough to allow four rounds of mostly synchronous divisions. For intermediate coupling, there is a sharp transition from \emph{D. melanogaster} to a set of small linear cysts with at most one branch. For strong coupling, division synchrony is recovered, and the final network returns to that of \emph{D. melanogaster}. 
See Fig. S3b for an analogous plot showing a more gradual effect of diffusion on network structure.

Together, these data give a physical understanding of how, by variation of parameters $v$ and $c$, the model generates a diverse set of naturally occurring topologies in invertebrate ovaries (Fig. \ref{fig:dros_oocyte_plus_species}).
In the Discussion, we consider how these principles for network formation -- combined with existing and future experimental data -- can further constrain where a given species sits in parameter space and thus the model-predicted effects of perturbations.

\section*{\label{sec:discussion}Discussion}

The generation of cellular networks by incomplete divisions is a gamete-production strategy observed commonly in metazoans, and the resulting network topologies differ considerably across species. 
Despite the evolutionary significance of these networks, we lack a quantitative framework for network formation and inter-species variability. 
Motivated by studies of invertebrate ovaries where such networks vary minimally within a species, we developed a model of network assembly based on replicating oscillators.
A regulatory pulse of strength $v$ promotes transient oscillations on network nodes (cells). 
Each oscillation causes an incomplete division that leaves the resulting cells coupled by diffusion with rate $c$. 
Asymmetric division ($\delta$) of the regulator allows heterogeneity between resulting cells.  
This model produces most known germline cyst topologies in invertebrates by variation of $v$, $c$, and $\delta$ (Fig. \ref{fig:cubesfigure}). 
For networks formed by $n$ rounds of synchronous divisions, our model explains how changing $v$ determines $n$, while for networks formed only by asynchronous division patterns, our model reveals how the interplay between $\delta$ and $c$ mediates such choreography. 
This analysis identifies where in parameter space natural networks reside and predicts parameter changes that transform one final network into another.

The network transitions predicted by our model can be tested in \emph{D. melanogaster} using genetic techniques. Deciding which genes to manipulate requires knowledge of where the corresponding gene products fit into our model. Fortunately, existing experimental evidence relates some specific proteins to our model parameters, particularly for the regulatory pulse $z$. 
For example, a protein called Bag-of-marbles (Bam) has a pulsed temporal profile and directly affects Cyclin A levels \cite{ShanmingChaoyi2017}, thus tuning the time-window of proliferation. Changing Bam expression has corresponding effects on network size, but this has been mostly characterized in males \cite{Insco2009}. Thus, future experimental measurements of Bam's time-course in females (wild-type and mutant) are required to establish its effect on model parameters like $v$.  
While there are no obvious individual genes controlling either cell-cell coupling or division asymmetry, mutations affecting a shared network-spanning organelle (called the fusome, Fig. S1c) cause \emph{D. melanogaster} to generate non-$2^{n}$ networks \cite{SnappIida2004, hts_mutant, McGrailHays1997, de_cuevas_alpha-spectrin_1996, yue_hu-li_1992} and plausibly correspond to changes in $c$ and $\delta$. Future work linking specific mutations to shifts in model parameters will enable model-informed experimental realization, in \emph{D. melanogaster} ovaries, of a diverse set of topologies -- including as-yet unreported networks predicted by the model like \emph{$\frac{3}{4}$-pint}, \emph{$\frac{5}{4}$-pint}, and \emph{one-short}.

A key additional test of the model relates not to the final network topology, but rather to the model-predicted trajectory by which the network forms. These tests require only the identification of network intermediates (e.g. by analyzing fixed germariums) and are particularly useful for species in which networks form by asynchronous divisions (e.g. \emph{N. vitripennis}, see Fig. S5d) \cite{EastinHuangFerree}. To illustrate how the model-predicted history can be tested, we consider \emph{C. perla}. Its primary variant has 12 cells and appears in three large regions of comparable size in our model parameter space. For two of these regions, the network forms by three rounds of synchronous divisions, then only four of eight cells divide in the last round (Fig. S2); in the other region, the two central cells skip the third round of divisions, then join all others in the last round (Fig. \ref{fig:buildlacewing}). The latter region matches the experimentally inferred history \cite{rousset_formation_1978}, and small parameter variations in that region -- but not in the other -- generate other reported \emph{C. perla} variants (Fig. S4). The future experimental reconstructions of such division sequences in other species will help to constrain where each wild-type species resides in parameter space, thus refining the model-predicted response to experimental perturbations.

Our work also establishes a foundation for studying networks that form less stereotypically, like the germline cysts in invertebrate testes \cite{diegmiller_fusome_2023, rasmussen_ultrastructural_1973}. In our model, sperm cysts' variability can be captured by modifying the effect of division on the network: compared to our previous assumption that divisions never modify existing edges, in sperm cysts a cell can instead divide into an existing network edge \cite{diegmiller_fusome_2023}. Whenever a new cell is generated, the potentially stochastic choice between these two modes results in  graph variability. This variability causes, for each point in parameter space, a deterministic final network to become a distribution of networks. Rather than transforming one graph into another, parameter changes shift the probability density in this space of graphs. Such a framework would enable the study of germline cyst formation even in mammals, where network fragmentation causes additional variability \cite{LeiLei2013}. From choanoflagellate colonies to the germline cysts in metazoans, cell networks' evolutionary importance demands a unified mathematical framework for their formation by oscillations, for which this work provides a foundation.


\section*{\label{sec:methods}Methods}

\subsection*{Modeling assumptions}
\label{appModel}

The assumptions underlying the presented model are based on experiments primarily in \emph{D. melanogaster}. Some assumptions may be violated in some other invertebrates; nonetheless, it is useful to construct a mathematical model based on a well-characterized species and, then, to test whether such a model is sufficiently rich to capture variation across species. 

The two most basic assumptions of the presented model are: 
(1) no cell death occurs during network formation,
(2) the intercellular bridges formed by incomplete division remain intact. 
Regarding (1), though the generation of an oocyte -- from this network of cells -- eventually requires the death of many cells in the network, this death does not occur during cyst formation, but rather much later \cite{foley_apoptosis_1998}. In fact, even in response to starvation of the fruit fly, germ cell death occurs only after the full network has formed \cite{drummond-barbosa_stem_2001}.
Regarding (2), in some invertebrates, some cells initially connected by ICBs disconnect (a process called fission or fragmentation) late in the network's formation or soon after the network formation \cite{buning_insect_1994}. Those cases are beyond the scope of this study.

In addition to the assumptions (1,2), the model contains five further rules (3-7):
(3) which intracellular factors cause cell division, (4) how one cell's state affects another's, (5) the incorporation of new cells into the network, (6) how mother-daughter asymmetries at divisions affect cell cycles, and (7) how the arrest of divisions is regulated. As outlined below, some of these assumptions -- (3, 5, 7) -- are more grounded in experimental data than others, like (4, 6).

For (3), we assume that the element that drives network growth is an oscillator since cell cycles -- oscillating levels of activated and inactivated cyclin-CDK complexes -- control when a cell divides \cite{DIEGMILLERREVIEW, JIALSOUS2021, HINNANT2017118, Tysoncdc2, TYSON2008R759, NOVAK1993101, NovakTyson2006}. The existing quantitative characterization of cell cycles in the germline indicates that they are complex, including not only S and M phases, but also gap phases \cite{HINNANT2017118}. Lacking measurements of cell cycle regulators from live-imaged cysts to constrain the parameters of such a model, we treat each cell as a simple two-dimensional relaxation oscillator -- motivated by cell cycles in early embryonic development of \emph{D. melanogaster} and \emph{X. laevis} \cite{strogatz2018nonlinear, DIEGMILLERREVIEW, QiongYang2013, FerrellYang2011, MATTINGLY2017743, DENEKE2016399, NOVAK1993101}.

For (4), we assume that the oscillations of one cell can affect those of another cell by diffusion over the network edges (ICBs) \cite{JIALSOUS2021, DOHERTY2021860, DIEGMILLERREVIEW}. Cell cycle coupling has long been hypothesized to be mediated by the fusome (Fig. S1c). Localization of a cyclin (in this case, active Cyclin A-CDK1) to the fusome could promote a trigger wave \cite{gelens_spatial_2014} along the fusome, if proteins promoting Cyclin A-CDK1 activation are present there \cite{LILLY200053}. Importantly, this coupling seems to require something like this reaction-diffusion scheme along the fusome \cite{LILLY200053}, as opposed to diffusion through shared cytoplasm alone \cite{SnappIida2004}.


For (5), we assume that, as observed in the female germline clusters of many invertebrates, each division introduces a node-edge pair, without altering existing edges. Motivating this assumption is experimental work which showed that the fusome (Fig. S1c), by anchoring one pole of the mitotic spindle, prohibits divisions from modifying existing network edges \cite{DIEGMILLERREVIEW, McGrailHays1997, StortoKing1989, TELFER1975223}. See the Discussion for a proposed model extension to handle the case of divisions into existing network edges.

For (6), asymmetric divisions can occur, leading to persistent differences in the oscillations of directly related cells. Though further experimental work is required, one hypothesized mediator of the asymmetry is unequal division of fusome volume, which plausibly leads to different effective concentrations of fusome-associated cell cycle regulators (like deubiquitinases) \cite{diegmiller_fusome_2023, de_cuevas_morphogenesis_1998, HUYNH2004R438}. Our assumption that the oscillating variables are copied at division is consistent with experimental observations. For example, Cyclin A, one of the key cyclins whose levels regulate the number of divisions, is diffuse in the cytoplasm at mitosis, so the difference in its concentration between the two resulting cells immediately after division is likely small \cite{LILLY200053}.

For (7), the number of divisions is likely regulated by a pulse of a cell cycle regulator  \cite{ShanmingChaoyi2017, Insco2009}. Existing experimental evidence indicates that a complex involving two proteins, called Bam and Ovarian tumor (Otu), acts to deubiquitinate Cyclin A. Furthermore, Bam is reported to have a ``pulsed"  concentration \cite{McKearin01121990, McKearinOhlstein1995, Insco2009} throughout cells of the network, increasing as the network starts to grow and decreasing as it stops, thus also changing Cyclin A degradation rates.

With respect to (7), the oscillator variables $x$, $y$ are affected by pulse $z$, but do not themselves appear in the differential equation for $z$. $z$ follows its own prescribed dynamics. In agreement with this treatment, the initiation of the pulse of Bam depends on factors external to the cell \cite{CHEN20031786}. Initial evidence also suggests a strong role for external factors, from some somatic cells surrounding the dividing cyst, in controlling the last divisions \cite{SHI2021840}. Thus, our treatment of the pulsed variable is in accordance with existing experimental evidence.

\subsection*{Mathematical details}
\label{appA}

\noindent
\\
\textbf{Number of model parameters} - There are five parameters of the single-cell relaxation oscillator. Three ($\epsilon,  \gamma, z$) appear directly in Eq. (\ref{eq2}), and two control the shape of $g(x)$ as explained in the next section. 
In total, the growing network model has eight free parameters: $z$ becomes a function of two parameters ($v$ and $t_p$, see below) and we also introduce the cell coupling $c$ and division asymmetry $\delta$. In this work, we are focused on the multicellular model and thus study variation of $v$, $c$, and $\delta$.

\noindent
\\
\textbf{Form of $g(x)$} - 
The piecewise-linear function $g(x)$ is given by

\begin{equation*}
  g(x) =
    \begin{cases}
 \:\:\:\:\:\, m x & \text{if \: $x\le \frac{1}{2}$}\\
       m(1 - x) & \text{if \: $\frac{1}{2} < x \le \frac{1 + a}{2}$}\\
       m(x - a) & \text{else}
    \end{cases}       
\end{equation*}
where $m$ and $a$ are positive shape parameters which we set to $m=2$, $a=\frac{1}{4}$ throughout. In general, $g(x)$ is constrained by requiring that $x$, $y$, and $y-x$ (activated, total, and deactivated cyclin-CDK) be non-negative for all times. 
In other words, the physical region bounded by the lines $\{x=0, y>0\}$ and $\{x>0, y=x\}$ must be a positively invariant set of the dynamics Eq. (\ref{eq2}). One then evaluates the derivative of $x$ and $y$ along these lines. For positive invariance, we trivially require $m>1$, and by asserting that $g(x)>x$ at the right cusp (a sufficient condition), one finds $\frac{m-1}{m+1} > a$ is also required.

\textbf{
\\Form of $z(t)$} - 
As depicted in Fig. \ref{fig:model_fig}b, $z(t)$ is modeled as a triangular pulse of the form
\begin{equation*}
z(t)=
    \begin{cases}
       z_0 + v t  & \text{if \: $0 \leq t \leq  t_p$}\\
       z_0 - v t  & \text{if \: $ t_p < t < 2 t_p$} \\
       z_0 & \text{else.}
    \end{cases}    
\end{equation*}
We refer to $v$ as the ``pulse strength" and $t_p$ as the ``pulse duration". The initial condition $z(t_0)\equiv z_0$ is set to $0$ throughout the main text. 


\noindent
\\
\textbf{Oscillatory regime} - 
From the intracellular dynamics Eq. \ref{eq2}, note that the $x$- and $y$-nullclines are defined by $y=g(x)$ and $y=\frac{z-x}{\gamma}$, respectively. When the nullclines intersect on the left ($L$) or right ($R$) branch of $g(x)$, the system tends to the associated stable fixed point. Conversely, when the nullclines intersect on the middle branch of $g(x)$, the fixed point is unstable and oscillatory behavior results. Thus, the corners of $g(x)$, $(x_L,y_L)=(\frac{1}{2}, \frac{m}{2})$ and $(x_R,y_R)=(\frac{1+a}{2}, \frac{m(1-a)}{2})$, can be used to define entrance and exit from the oscillatory region (annotated in Fig. \ref{fig:model_fig}a).

Recall that $z(t)$ controls entrance and exit from the oscillatory region by translating the $y$-nullcline (Fig. \ref{fig:model_fig}a). 
In the absence of cell coupling, the oscillatory regime is given by $z_L < z < z_R$, where 
\begin{equation*}
z_L = x_L + \gamma y_L = \frac{1 + \gamma m}{2}
\end{equation*}
and
\begin{equation*}
z_R = x_R + \gamma y_R = \frac{1 + a + \gamma m(1-a)}{2}.
\end{equation*}
Note that $z_R = z_L + \frac{a}{2}(1 - \gamma m)$.
Finally, in the limit $\gamma \rightarrow 0$ the oscillatory regime is simply $\frac{1}{2} < z(t) < \frac{1+a}{2}$.

\noindent
\\
\textbf{Cycle detection} - 
For the minimal single-cell oscillator considered here, we define division events based on completed periods of its limit cycle. 
To detect these periods, we use repeated threshold crossing of the fast variable $x$. 
A complete cycle occurs when $x$ crosses a threshold $x_T$ from the left, then the right, then from the left once more in sequence. 
We use $x_T=\frac{2+a}{4}$ which is the center of the oscillation region defined by $g(x)$ (depicted by the vertical dashed lines in Fig. \ref{fig:model_fig}a). 

\noindent
\\
\textbf{Heuristic for number of cycles caused by the regulatory pulse} - 
As depicted in Fig. \ref{fig:twoparameterfig}a, the number of full oscillations (cycles) of an independent cell corresponds to an integral over the inverse period of the oscillation ($T(z)$; see SI Text for an analytic approximation) at changing values of $z$. 
To that end, define 
\begin{equation*}
S(v)=\frac{1}{v} \int_{z_L}^{\min{(z_R, v t_p)}} \frac{1}{T(z)} dz. 
\end{equation*}
which counts the accumulated phase for one-half of a symmetric pulse of strength $v$. Integer values of $S(v)$ correspond to full cycles that occur during the rising segment of the pulse. 
Note that the limits of integration account for the case of the pulse crossing the upper boundary of the oscillatory region $v t_p > z_R$.

A heuristic for the number of cycles of an independent oscillator, and thus the size of the network in the case of symmetric divisions ($\delta=0$), is then simply
\begin{equation*}
n(v) =
    \begin{cases}
       0   & \text{if \: $v t_p \le z_L$}\\
       \lfloor 2 S(v) \rfloor  & \text{if \: $z_L < v t_p \le z_R$} \\
       2 \lfloor S(v)  \rfloor        & \text{else.}
    \end{cases}    
\end{equation*}
If the pulse $z(t)$ never enters the oscillatory zone, then by definition there can be no cycles. If $z(t)$ enters but does not exceed the upper boundary $z_R$, we do not need to consider the ``resetting" of the phase that occurs when it overshoots (i.e. $v t_p > z_R$). Note that we take the floor $\lfloor \cdot \rfloor$ because the network only grows when full mitotic cycles are completed. 

\subsection*{Computational details}

\noindent
\\
\textbf{Simulating network formation} - 
We developed code to simulate the network formation process outlined in Fig. \ref{fig:model_fig}. 
The code is written in Python 3.8 and relies on various libraries including NumPy and SciPy. 

Our approach includes vectorized routines to numerically solve the (growing) system of ODEs Eq. (\ref{eq:multicell}). 
In brief, given an initial network configuration (e.g. a single cell in state $\mathbf{x}^{(0)}$ at time $t_0$), we integrate the system for a fixed time window $\Delta t$.
We then inspect the full trajectory for completed intracellular oscillation events. There are two cases:
\begin{enumerate}
    \item No completed oscillations are detected. 
    \item At least one oscillation is detected. Let $t^*$ denote the completion of the first oscillation which has not yet been detected (note $t_0 \le t^* \le t_0 + \Delta t$). We discard the portion of the trajectory after $t^*$, and expand the graph by initializing a daughter cell (introducing $N$ additional dynamic variables). The expanded network at time $t^*$ is treated as an initial condition for the next iteration. 
\end{enumerate}
We repeat this process of incrementally extending the network trajectory by $\Delta t$ and inspecting for oscillations until the network stops growing, which is detectable by $\mathbf{x}_a \rightarrow \mathbf{0}$ for all cells $a \in \{1, \ldots, M\}$. 
The code includes numerous other functionalities, such as graph isomorphism checks to compare the growing network against a library of named graphs (e.g. those of Fig. \ref{fig:dros_oocyte_plus_species}c-f).

\noindent
\\
\textbf{Parameter sets} - 
To identify natural networks produced by the model, we first simulated network formation for each point in a three-dimensional grid of 
$100 \times 160 \times 181$ parameter sets $(v, c, \delta) \in [0.005, 0.017] \times [0, 8] \times [-0.09, 0.09]$. 
Consolidated data from this grid appear in Fig. \ref{fig:cubesfigure}b of the main text (explained further in the next subsection).
The large dataset also forms the basis for individual network trajectories and one-dimensional parameter scans shown in other figures. 

Specific parameter values used to generate different items in the text are listed in Table S2.
Parameter values used throughout the text are: $a=\frac{1}{4}$, $m=2$, $\gamma=10^{-2}$, $\epsilon=10^{-2}$.
Furthermore, all networks were initialized as a single cell in state $(x^{(0)}, y^{(0)}, z^{(0)})=(0,0,0)$. 

\noindent
\\
\textbf{Space of naturally occurring networks produced by varying parameters} - 
Our goal is to identify region(s) in the grid introduced above that produce networks which are isomorphic to one of the reported naturally occurring ones. 
To that end, let $\{\mathbf{A}_1, \ldots, \mathbf{A}_S\}$ denote a set of $S$ target networks (e.g. those in Fig. \ref{fig:dros_oocyte_plus_species}c-f), and let $\mathbf{A}^*_{i,j,k}$ be the final network produced at point ($v_i$, $c_j$, $\delta_k$) in parameter space. 
From these, we construct a tensor $\Lambda_{i,j,k}$ where $\Lambda_{i,j,k}=p$ if $\mathbf{A}^*_{i,j,k}$ is graph isomorphic to $\mathbf{A}_p$, and $\Lambda_{i,j,k}=0$ if it does not match any target network.

For each target network $\mathbf{A}_p$, we then use regionprops3 (MATLAB, voxel connectivity of 26) to identify the connected component(s) in our tensor (this is how we define ``regions" of parameter space). 
We disregard any component which contains only one voxel. 
To check if two regions generating distinct target networks ($\mathbf{A}_p$, $\mathbf{A}_q$) are in contact, we again use regionprops3 (as above). 

Though the analysis outlined above often identifies more than one region that generates a given target network $\mathbf{A}_p$, in Fig. \ref{fig:cubesfigure}b we plot a single point representing a particular region. 
See Fig. S4 for an example of how the underlying regions appear for \emph{D. melanogaster} and \emph{C. perla}.
In most cases, including that of \emph{D. melanogaster}, the point corresponds to the centroid of the largest corresponding region. 
For each of the networks contacting \emph{D. melanogaster} in parameter space (see Fig. \ref{fig:cubesfigure}a), we specifically choose a representative region contacting \emph{D. melanogaster}.
Finally, the size of each point in Fig. \ref{fig:cubesfigure}b corresponds logarithmically to the voxel count of the region it represents.

\begin{acknowledgments}
We thank Jasmin Imran Alsous, Rocky Diegmiller, Trudi Sch{\"u}pbach, John Tyson, Elizabeth Ables, Allison Beachum, and Steven Strogatz for helpful discussion related to the model and/or manuscript, and Lucy Reading-Ikkanda for illustrations. 
M.S. acknowledges the support of NSERC MSFSS and Mitacs GRA. 
\end{acknowledgments}

\begin{center}
    \textbf{Supplementary Information}
\end{center}
The following Supplementary Information is available for this paper:
\begin{itemize}
    \item Supplementary Text. 
    \item Supplementary Tables S1-S2.
    \item Supplementary Figures S1-S5.
\end{itemize}

\bibliography{references.bib}

\end{document}


\maketitle

This supplementary file contains:
\begin{itemize}
\item Supplementary text
\item Supplementary Tables S1 - S2
\item Supplementary Figures S1 - S5
\end{itemize}

\newpage

\subsubsection*{Heuristic calculations}

\noindent
\\
\textbf{Heuristic for the oscillation period $T(z)$} - 
The limit cycle associated with the relaxation oscillator is approximated by clockwise motion on a parallelogram in the $xy$-plane (see Fig. 2a). From the upper left, its corners are:


\begin{align*} 
\label{eqS3}
(x_L,y_L) &=\left(\frac{1}{2}, \frac{m}{2} \right) \\
(x_{R'},y_{R'}) &=\left(\frac{1}{2} + a, \frac{m}{2} \right) \\
(x_R,y_R) &=\left(\frac{1+a}{2}, \frac{m(1-a)}{2} \right) \\
(x_{L'},y_{L'}) &=\left(\frac{1-a}{2}, \frac{m(1-a)}{2} \right).
\end{align*}

The period $T$ for a fixed value of $z$ (i.e. $T(z)$) involves fast and slow contributions from motion along the parallelogram edges. 
The fast contributions -- which we will neglect -- come from horizontal motion of $x$ jumping between branches of $g(x)$ (i.e. $L\rightarrow R'$ and $R\rightarrow L'$), while the slow contributions arise from crawling up or down a branch of $g(x)$ (i.e. $L'\rightarrow L$ and $R'\rightarrow R$ ) in between these fast jumps. 

Note that $T(z)$ will be dominated by the slow contributions outlined above. We compute these as two separate cases below, as they involve different branches of $g(x)$. To proceed, we leverage $\epsilon \ll 1$ to make a quasi-steady state approximation for the fast variable $x$ such that $y \approx g(x)$. 

\textbf{Case A}: $L'$ to $L$ -- Here $g(x)=m/2$ and so $x^*=y/m$. 
Substituting this in Eq. (2) and solving the remaining ODE in $y(t)$ for initial condition $y_L'$ gives 
\begin{equation}
\label{eqS4}
y_A(t)=\frac{2 m}{z_L} \left(z-e^{-2 m z_L t} (z - (1 - a)z_L)\right)
\end{equation}
Then, to find the contribution to the period, solve $y_A(T_A)=y_L$ to find:
\begin{equation}
\label{eqS5}
T_A = \frac{m}{2 z_L} \ln \left(1 + \frac{a z_L}{z - z_L}\right).
\end{equation}

\textbf{Case B}: $R'$ to $R$ -- On the right branch we instead have $g(x)=m(x-a)$, so that $x^*=y/m + a$. 
Repeating the steps above leads to
\begin{equation}
\label{eqS6}
T_B = \frac{m}{2 z_L} \ln \left(1 + \frac{a z_L }{z_R - z}\right).
\end{equation}

Summing these contributions above gives the limit cycle period heuristic
\begin{equation}
\label{eqS7}
T = 
\frac{m}{2 z_L} \ln \left(\left(1 + \frac{a z_L}{z - z_L}\right)\left(1 + \frac{a z_L}{z_R - z}\right)\right)
\end{equation}
Note that this simplifies to $T = m \ln \left(\left(1 + \frac{a}{2 z - 1}\right)\left(1 + \frac{a}{1+a - 2 z}\right)\right)$ in the $\gamma \rightarrow 0$ limit.
Finally, while the heuristics in this section rely on $\epsilon \rightarrow 0$ (i.e. the quasi-steady state approximation), they are reasonably accurate (Fig. 5a) at $\epsilon=10^{-2}$ which we use throughout the text. 

\noindent
\\
\textbf{Heuristic for the range of $c$ leading to oscillator death} - 
To probe the effect of coupling shown in Fig. 5d, we consider two coupled oscillators. This derivation is only meant to show that the non-monotonic effect of diffusion arises naturally in our model, requiring neither dynamics for $z$ nor complex network topologies. For simplicity, we assume that the oscillators have static $z$ values and are initially desynchronized: one oscillator (1) on the right branch of the cycle, the other (2) on the left.
The $x$ dynamics are as follows:
\begin{equation}
\label{eqS8}
\begin{split}
 \frac{d x_1}{dt}  = 
    \frac{1}{\epsilon}\Bigl((y_1 - m x_1) 
    &+ c \epsilon (x_2 - x_1)\Bigr) \\
 \frac{d x_2}{dt}  = 
    \frac{1}{\epsilon}\Bigl((y_2 - m (x_2 - a)) 
    &+ c \epsilon (x_1 - x_2)\Bigr)
\end{split}
\end{equation}
where we have substituted in the functional form of $g(x)$ for $x\le \frac{1}{2}$ (1) and $x \ge \frac{1 + a}{2}$  (2).

For $0 < \epsilon \ll 1$, Eq. (\ref{eqS8}) reduces to the approximate relations below:
\begin{equation}
\label{eqS9}
\begin{split}
 (y_1 - m x_1) 
    &+ c \epsilon (x_2 - x_1) \approx 0 \\
 (y_2 - m (x_2 - a)) 
    &+ c \epsilon (x_1 - x_2) \approx 0
\end{split}
\end{equation}

The two relations above (Eq. (\ref{eqS9})) approximately constrain both $x_1$ and $x_2$. On the other hand, $y_1$ and $y_2$ evolve slowly relative to $x_1$ and $x_2$. That slow evolution can be written as a difference and sum, respectively:
\begin{equation}
\label{eqS10}
\begin{split}
\frac{d (y_1 - y_2)}{dt}  = 
    (z_1 - z_2) - (x_1 - x_2) - \gamma (y_1 - y_2)  
    &+ 2 c (y_2 - y_1) \\
\frac{d (y_1 + y_2)}{dt}  = 
    (z_1 + z_2) - (x_1 + x_2) - \gamma (y_1 + y_2)
\end{split}
\end{equation}
Substitution of the constraints (Eq. (\ref{eqS9})) on $x_1$ and $x_2$ into Eq. (\ref{eqS10}) produces:
\begin{equation}
\label{eqS11}
\begin{split}
\frac{d (y_1 - y_2)}{dt}  = 
    (z_1 - z_2) + \Bigl(\frac{a m - y_1 + y_2}{2 c \epsilon + m}\Bigr)  - \gamma (y_1 - y_2)  
    &+ 2 c (y_2 - y_1) \\
\frac{d (y_1 + y_2)}{dt}  = 
    (z_1 + z_2) - \Bigl(\frac{a m + y_1 + y_2}{m}\Bigr) - \gamma (y_1 + y_2)
\end{split}
\end{equation}
To see if oscillator death emerges for some values of $c$, we can then solve for the ``fixed point" $(y_1 - y_2, y_1 + y_2)_\textrm{FP}$ :
\begin{equation}
\label{eqS12}
(y_1 - y_2, y_1 + y_2)_\textrm{FP} = \Bigl(\frac{a m + (2 c \epsilon + m) (z_1 - z_2)}{1+(2 c + \gamma)(2 c \epsilon + m)}, \frac{m(-a+z_1+z_2)}{1 + \gamma m}\Bigr)
\end{equation}
This groupwise fixed point (i.e. oscillator death) is trivially stable when it exists, which is true when $x_1 \le \frac{1}{2}$ and $x_2 \ge \frac{1 + a}{2}$ (since we have replaced $g(x)$ by its functional form on the left (1) and right (2) branch, see Eq. (\ref{eqS8})). 
To verify that the fixed point $(y_1 - y_2, y_1 + y_2)_\textrm{FP}$ satisfies the specified inequalities in $x_1$ and $x_2$, we first convert $(y_1 - y_2, y_1 + y_2)_\textrm{FP}$ to equations for $y_{1,\textrm{FP}}$ and $y_{2,\textrm{FP}}$, then calculate $x_{1,\textrm{FP}}$ and $x_{2,\textrm{FP}}$ using Eq. (\ref{eqS9}).
The result is that, for oscillator death to occur, the following constraints on parameters must be satisfied:
\begin{equation}
\label{eqS13}
\begin{split}
    x_{1,\textrm{FP}} = \frac{a c (-1 + \epsilon \gamma (2 c + \gamma)) m + (1 + \gamma m) z_1 + 
   (2 c^{2} \epsilon + 
   c (\epsilon \gamma + m)) (z_1 + z_2)}{(1 + 
     \gamma m) (1 + (2 c + \gamma) (2 c \epsilon + m))}
     &<\frac{1}{2} \\
    x_{2,\textrm{FP}} = x_{1,\textrm{FP}} +\frac{a (2 c + \gamma) m - (z_1 - z_2)}{1 + (2 c + \gamma) (2 c \epsilon + m)}>\frac{1+a}{2}
\end{split}
\end{equation}
The $c$ values which jointly satisfy Eq. (\ref{eqS13}) cause the two coupled oscillators to exhibit oscillator death if they are initialized on the opposite branches of the limit cycle. 

To check whether such a range in $c$ exists, we solve for the two $c$ values satisfying $x_{1,\textrm{FP}} = \frac{1}{2}$ ($c_{1}^{-}$ and $c_{1}^{+}$) and the two $c$ values satisfying $x_{2,\textrm{FP}} = \frac{1+a}{2}$ ($c_{2}^{-}$ and $c_{2}^{+}$). Suppose that all these roots are positive real numbers. If $c_{1}^{+}$ -- greater than $c_{1}^{-}$ by definition -- is greater than $c_{2}^{-}$, then there exists a range in $c$ for which oscillator death (a groupwise fixed point) occurs. 
For example, for our typical values of $a$, $m$, $\gamma$, $\epsilon$ (see Methods) and $z_1 = z_2 = \frac{3}{5}$, $c_{1}^{-}$ and $c_{1}^{+}$ are approximately 0.7 and 37.2, respectively, while $c_{2}^{-}$ and $c_{2}^{+}$ are approximately 0.1 and 264.2. Thus, both $c_{1}^{-}$ and $c_{1}^{+}$ lie within the interval $\left[  c_{2}^{-} , c_{2}^{+}  \right]$. In that case, oscillator death occurs for $c$ within the interval $\left[  c_{1}^{-} , c_{1}^{+}  \right]$. 
For other parameter choices (including the static $z_1$, $z_2$), the interval for $c$ in which oscillator death occurs can be straightforwardly calculated.

\newpage

\subsubsection*{Generalizations of the model}

\noindent
\\
\textbf{History-dependent asymmetry $\delta$} - 
In the main text, we assume that $\delta$ -- the model parameter that controls how asymmetrically $z$ is distributed between cells following a division event -- is fixed. 
We consider here an extension in which $\delta$ varies based on properties of the mother cell. Specifically, we let $\delta$ be a function of $k$, the number of times the mother cell has previously divided,
\begin{equation}
\label{eqS1}
    \tilde \delta(k) = \delta + \frac{\beta}{1+k},
\end{equation}
where $\beta$ tunes the strength of the history-dependence. 
Note that when $\beta=0$, we recover the fixed asymmetry used in the main text. 

With this mild generalization, the model additionally accounts for germline networks from female \emph{Dactylobiotus parthenogeneticus} (water bear) and \emph{Linepithema humile} (Argentine ant) (see Table \ref{tab:parameters} for examples). 

\noindent
\\
\textbf{Anisotropic diffusion} - 
In general, the rate of cell-cell diffusion $c$ may be different for each component of the intracellular state $\mathbf{x}\in\mathbb{R}^N$. 
To account for this in the model, define the corresponding vector $\mathbf{c}\in\mathbb{R}^N$ of diffusion coefficients and let $\mathbf{C}=diag(\mathbf{c})$. 
We also introduce explicitly the graph Laplacian $\mathbf{L}=\mathbf{D}-\mathbf{A}$ associated with a network adjacency matrix $\mathbf{A}$,   where the degree matrix $\mathbf{D}$ is diagonal with entries given by the row-sums of $\mathbf{A}$.
Finally, define the $M \times N$ matrix $\mathbf{X}$ where the $k^{th}$ row represents the state of the $k^{th}$ cell. 
Then the anisotropic diffusion dynamics on the network of cells is 
\begin{equation}
\label{eqS2}
\frac{d\mathbf{X}}{dt}=\left[\mathbf{f}\left(\mathbf{x}_1\right)\cdots\mathbf{f}\left(\mathbf{x}_M\right)\right]^T-\mathbf{LXC},
\end{equation}
where we have scaled each column $i$ of $\mathbf{LX}$ by the corresponding components of $\mathbf{c}$. 
The case of isotropic diffusion along each intracellular component reduces to a matrix representation of Eq. (3) in the main text.

\newpage

\begin{table*}\centering
\caption{Overview of model terms and parameters.}

\begin{tabular}{ccl}
Expression & Dimension & Description \\
\midrule
$M$         & scalar           & 
    Number of cells (nodes) in the network. \\
$N$         & scalar           & 
    Number of intracellular state variables; $N=2$. \\
$x$       & scalar           &
    Oscillatory variable with fast timescale.\\
    ~ & ~ & (analogue of activated cyclin-CDK).\\
$y$       & scalar           &
    Oscillatory variable with slow timescale.\\
    ~ & ~ & (analogue of total cyclin-CDK).\\
$\mathbf{x}$ & $N$          &
    State vector of a given cell; $\mathbf{x}=(x,y)$.\\
$\mathbf{f}(\mathbf{x})$ & $\mathbf{f}:\mathbb{R}^N \rightarrow \mathbb{R}^N$ & 
    Dynamical system representing cell cycle.\\
$\mathbf{h}(t)$ & $\mathbf{h}:\mathbb{R} \rightarrow \mathbb{R}^N$ & 
    Time-dependent cell cycle control; $\mathbf{h}=(0, z(t))$.\\
$\epsilon$  & scalar           & 
    Controls timescale of $x$ dynamics relative $y$. \\
$\gamma$    & scalar           & 
    Controls slope of $y$-nullcline. \\
$g(x)$      & $g:\mathbb{R} \rightarrow \mathbb{R}$ &          
    Encapsulates nonlinear regulation of $x$. \\
$z$       & scalar           & 
    Controls whether cell is in oscillatory regime. \\
    ~ & ~ & See definition of $\mathbf{h}(t)$.\\
$v$         & scalar           & 
    Magnitude of slope of triangular pulse of $z$. \\
$t_p$       & scalar           & 
    Midpoint of the triangular pulse of $z$. \\
$\mathbf{A}$ & $M \times M$   &
    Adjacency matrix for the multicell network. \\
$c$          & scalar             & 
    Diffusion rate of intracellular $x, y$ across edge. \\
$\delta$   & scalar             &  
    Asymmetry of $z$ for mother/daughter. \\
\bottomrule
\end{tabular}
\label{tab:ExpressionsVariables}
\end{table*}

\newpage

\begin{table}[h]
\centering
\caption{Parameter values used to generate naturally occurring networks}
\tableshift
\begin{tblr}{
  colspec={|Q| Q Q Q Q Q|},
  rows={halign=r},
  column{1}={halign=l},
  column{2}={halign=l},
  column{3}={halign=c},
  column{4}={halign=c},
  column{5}={halign=c},
  column{6}={halign=c},
  row{1}={halign=c},
}
\hline
Text item & $v$ & $c$ & $\delta$ & $\beta$ & $t_p$ \\[0.5ex] 
\hline 
\hline 
Fig. 3 (\emph{C. perla}; R1)& 
    $0.008375$
    & $0.3$
    & $-0.004$
    & $0$
    & $75$ \\ 

\hline 
Fig. 5a & 
    varies
    & $0$ 
    & $0$
    & $0$
    & $82$ \\ 

\hline 
Fig. 5c & 
    $0.01025$
    & varies
    & $0.004$
    & $0$
    & $82$ \\ 

\hline 
Network: \emph{D. parthenogeneticus} & 
    $0.0045$
    & $8.0$
    & $0.2$
    & $-0.1$
    & $150$ \\ 

\hline 
Network: \emph{L. humile} & 
    $0.0051$
    & $0.3$
    & $0.008$
    & $-0.004$
    & $150$ \\ 

\hline 
Fig. S2 (\emph{C. perla}; R2) & 
    $0.01125$
    & $2.8$
    & $0.01$
    & $0$
    & $60$ \\ 

\hline
Fig. S3a & 
    varies
    & $2.39196$
    & $0.007$
    & $0$
    & $82$ \\ 

\hline
Fig. S3b & 
    $0.0092625$
    & varies
    & $-0.013$
    & $0$
    & $82$ \\ 

\hline
Fig. S5a (5/4-\emph{pint}) & 
    $0.00826$
    & $6.1$
    & $0.016$
    & $0$
    & $82$ \\ 

\hline
Fig. S5b (3/4-\emph{pint}) & 
    $0.0104$
    & $6.1$
    & $0.023$
    & $0$
    & $82$ \\ 

\hline
Fig. S5c (\emph{one-short}) & 
    $0.0103$
    & $5.3$
    & $-0.02$
    & $0$
    & $82$ \\ 

\hline
Fig. S5d (jewel wasp)& 
    $0.00766$
    & $0.151$
    & $0.026$
    & $0$
    & $82$ \\ 

 \hline
\end{tblr}
\tableshift
\label{tab:parameters}
\end{table}

\newpage

\begin{figure}[!ht]
\centering\includegraphics[width=0.7\linewidth]{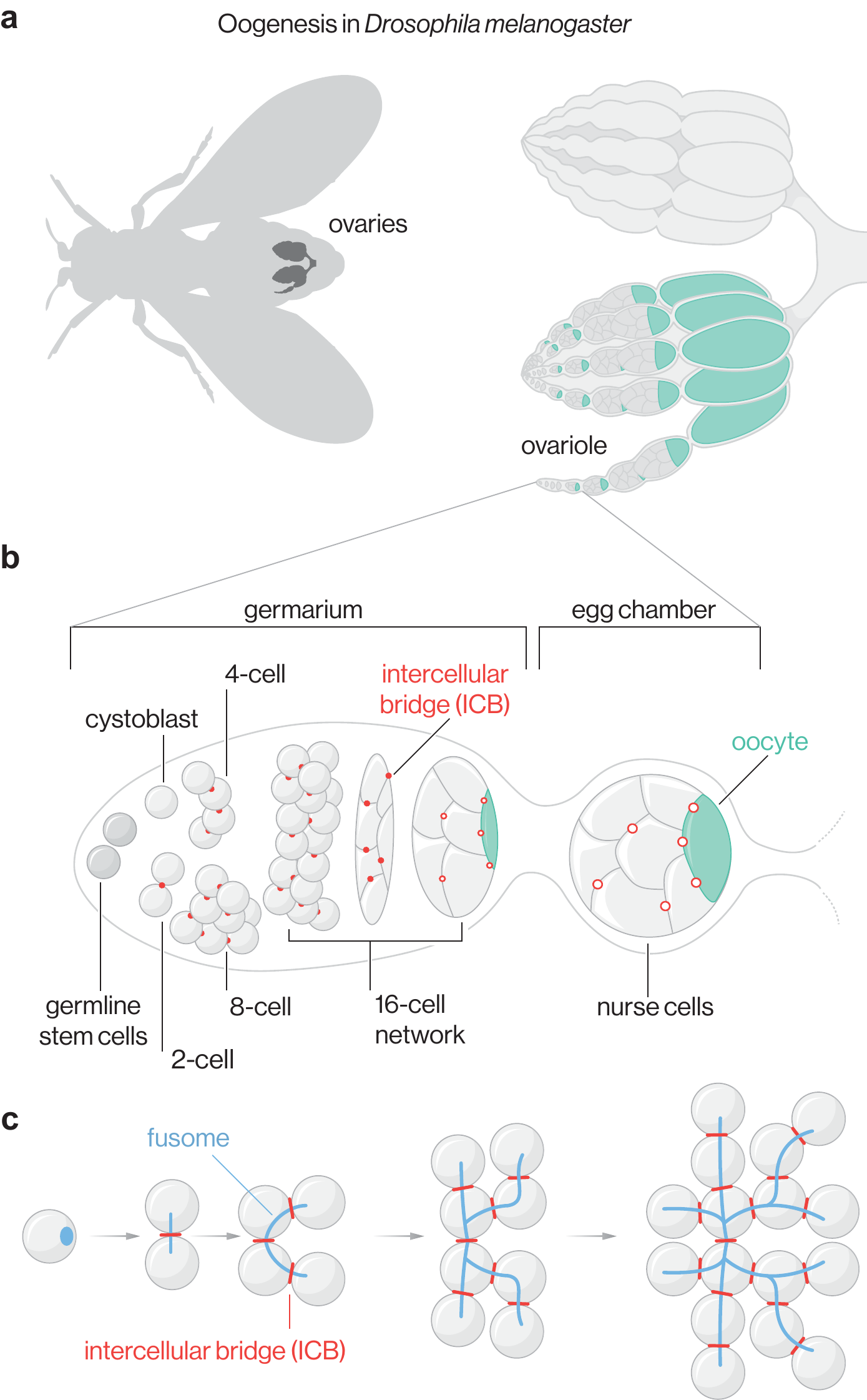}
\caption{\textbf{Germline cysts of female \emph{D. melanogaster}.} 
\textbf{a}, Ovaries are located in the female’s abdomen and contain multiple ovarioles. 
\textbf{b}, Ovarioles serve as assembly lines for germline cysts and associated oocytes. 
The germarium lies at the anterior end of the ovariole and houses the germline stem cells that undergo asymmetric division to produce cystoblasts. 
These cystoblasts -- which represent the initial founding cell in our model of network formation (see e.g. Fig. 3) -- undergo several incomplete divisions to produce a small network of cells connected via intercellular bridges from which oocytes are selected. 
In \emph{D. melanogaster}, the invariant 16-cell branched topology (see Fig. 1b) is produced through four rounds of incomplete divisions that are synchronous and highly stereotypic.
After network assembly, one of the central cells is specified as the oocyte while the other 15 become nurse cells and the cyst is then packaged in the egg chamber.
\textbf{c}, In addition to being cytoplasmically coupled throughout network assembly, all cells are connected by a shared organelle called the fusome that spans across each intercellular bridge. 
}
\label{fig:oogenesis}
\end{figure}

\pagebreak


\begin{figure}[!ht]
\centering\includegraphics[width=1.0\linewidth]{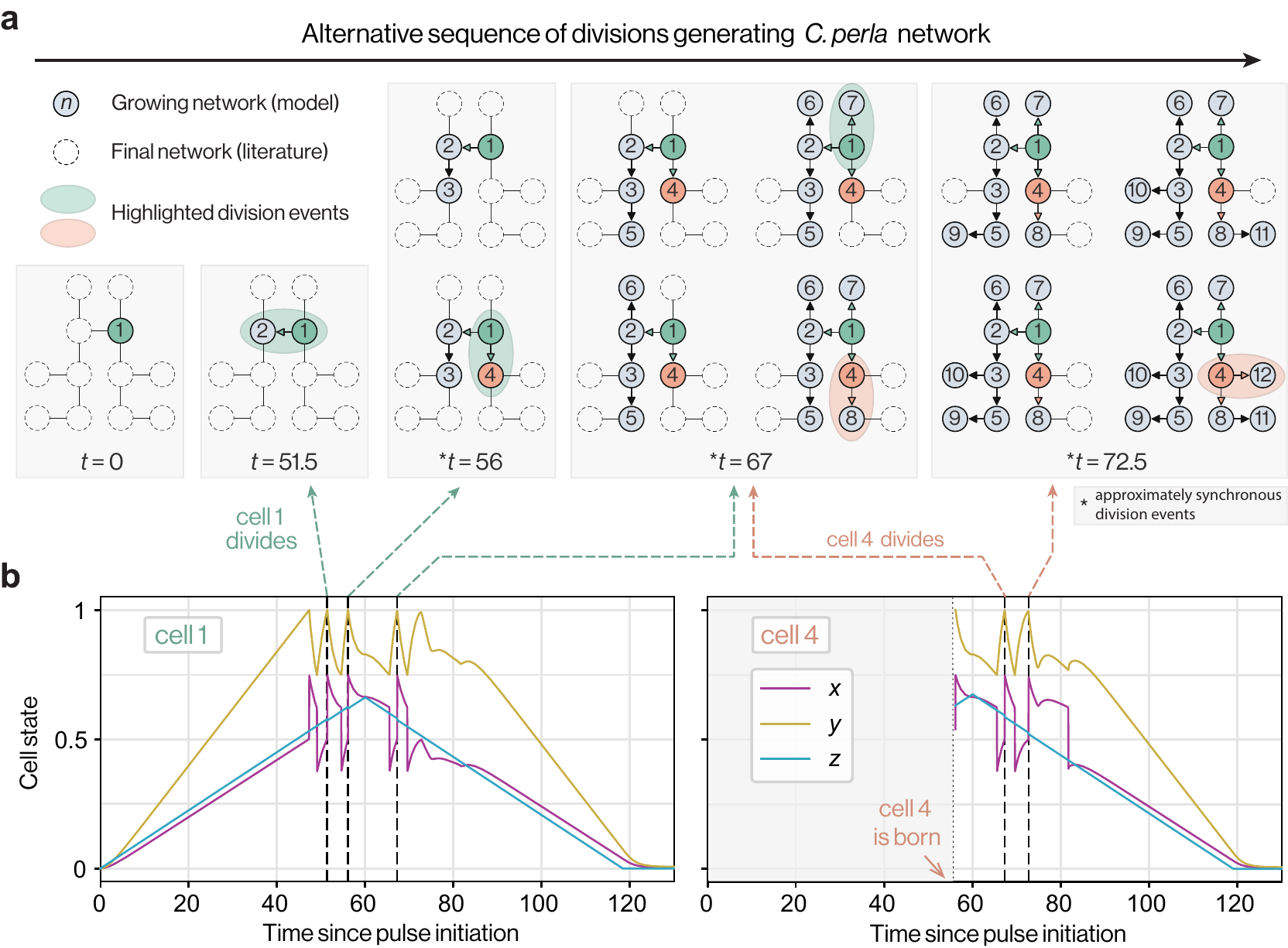}
\caption{\textbf{Alternative path to generating \emph{C. perla} network.} 
Panel layout closely follows Fig. 3 in the main text. 
The same final network as in Fig. 3 (i.e. the primary \emph{C. perla} germline cyst) can be assembled by a qualitatively distinct division sequence for a different set of model parameters.
}
\label{fig:S2}
\end{figure}

\begin{figure}[!ht]
\centering\includegraphics[width=1.0\linewidth]{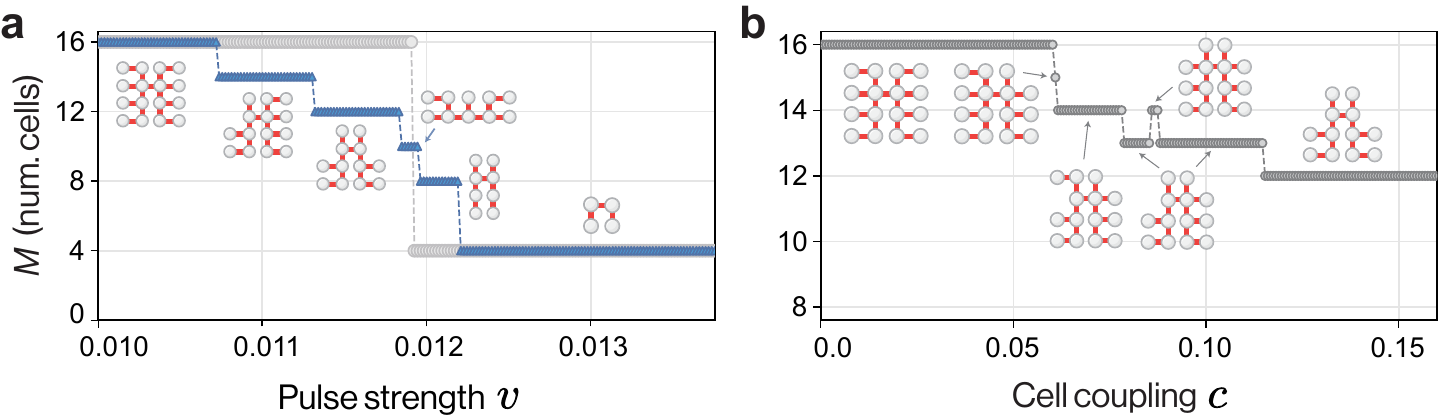}
\caption{\textbf{Supplemental data for Fig. 5.} 
\textbf{a}, Final number of cells $M$ in networks generated by asymmetric divisions with intermediate $c$ (blue triangles). 
Non-$2^{n}$ networks emerge for ranges in $v$ corresponding to transitions in Fig. 5a.
Gray points in \textbf{a} correspond to Fig. 5a of the main text. 
\textbf{b}, Size $M$ of networks generated by weakly negative asymmetric divisions ($-1 \ll \delta < 0$). 
By small changes in $c$, the final network transitions from the primary \emph{C. perla} variant (right) directly to its secondary variant and then to minor variants. 
}
\label{fig:S3}
\end{figure}

\begin{figure}[!ht]
\centering\includegraphics[width=1.0\linewidth]{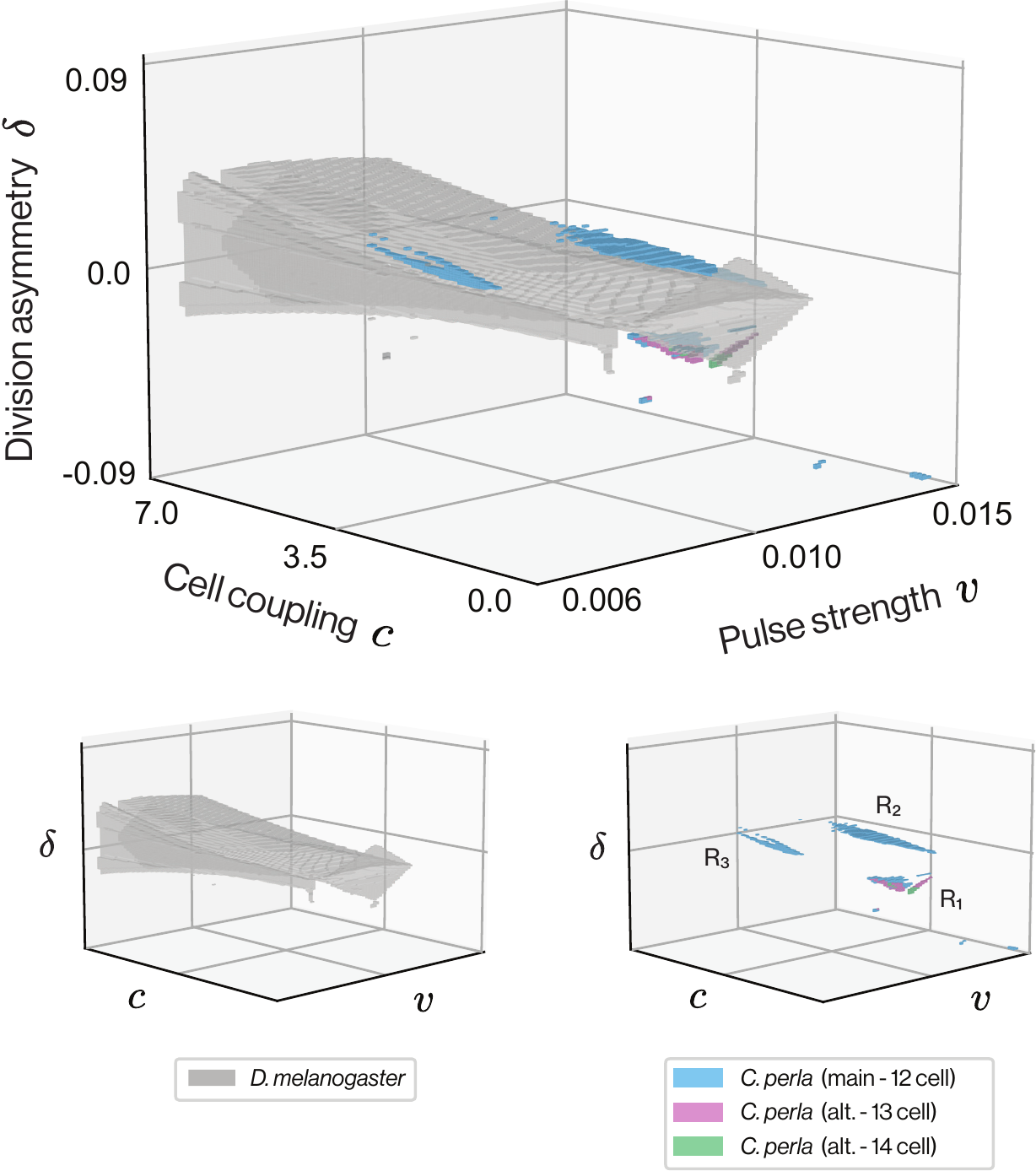}
\caption{\textbf{Regions of parameter space that produce ovarian germline cysts of \emph{D. melanogaster} and \emph{C. perla}.} 
Points in parameter space that produce four different natural networks (distinguished by color). Points which generate \emph{D. melanogaster} are shown in the bottom left, and \emph{C. perla} in the bottom right. The upper image overlays the lower images. Distinct regions that generate \emph{C. perla} networks are denoted by $R_1, R_2, R_3$. Of these regions, only $R_1$ produces secondary and tertiary variants of the primary \emph{C. perla} network. 
}
\label{fig:S4}
\end{figure}

\begin{figure}[!ht]
\centering\includegraphics[width=0.7\linewidth]{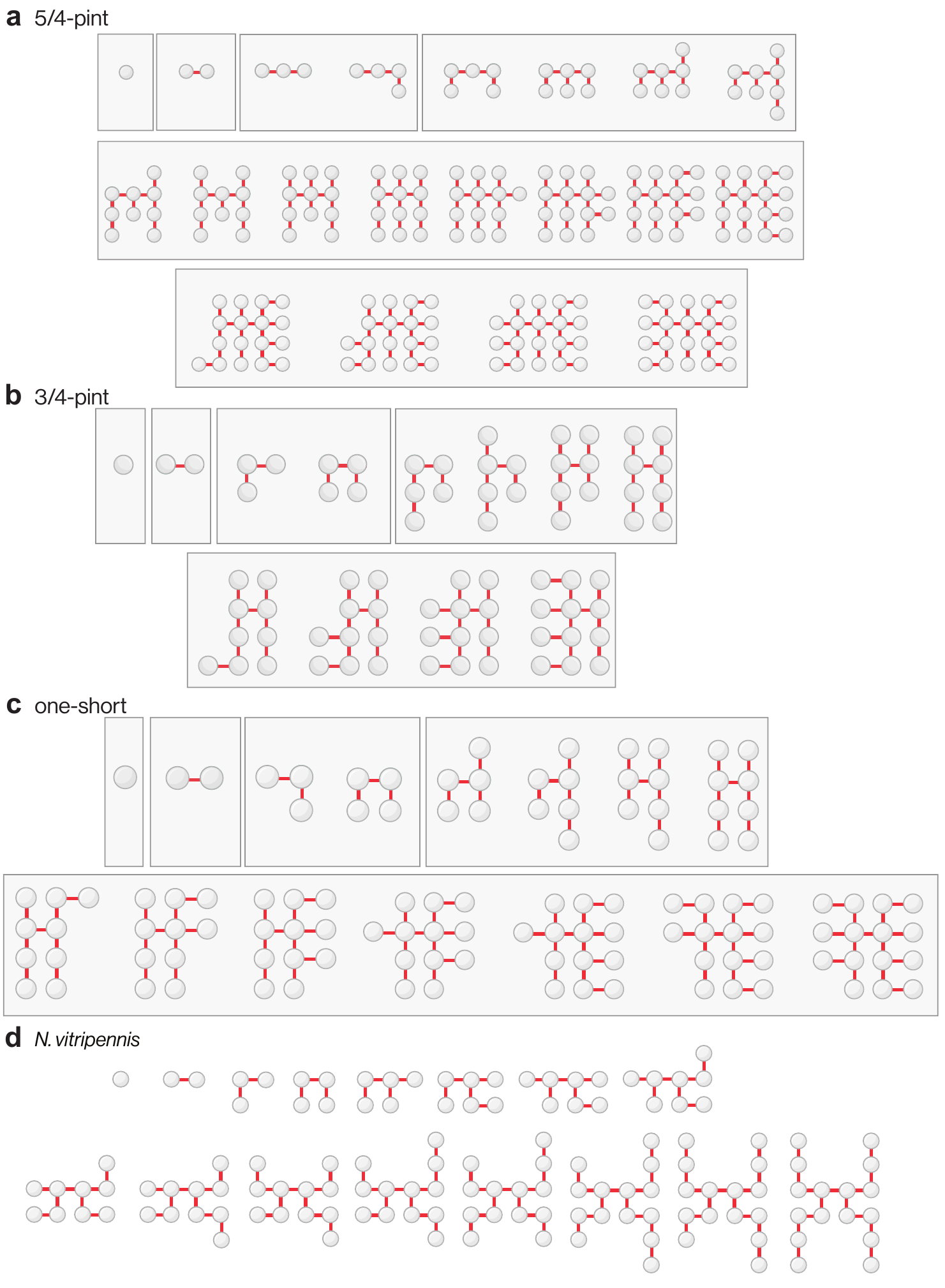}
\caption{\textbf{Example model-predicted sequences of divisions producing specific final network topologies.}
Approximately synchronous divisions are grouped by a gray bounding box: any two graphs in the same box must be generated within 0.1 simulation time units of each other. 
For \textbf{d}, all divisions after the 4-cell stage are quite asynchronous, so there exists no clear grouping of divisions.
\textbf{a}, \emph{$\frac{5}{4}$-pint}
\textbf{b}, \emph{$\frac{3}{4}$-pint}
\textbf{c}, \emph{one-short}
\textbf{d}, \emph{Nasonia vitripennis} (jewel wasp).
}
\label{fig:S5}
\end{figure}